# IDENTIFICATION OF LOCAL ALFVÉN WAVE RESONANCES WITH REFLECTOMETRY AS A DIAGNOSTIC TOOL IN TOKAMAKS

A.G.Elfimov[1], R. M.O. Galvão[1], L. F. Ruchko[1], and M.E.C.Manso[2]

[1]*Instituto de Física, Universidade de São Paulo, 05315-970, SP, Brazil*

[2]*Centro Fusão Nuclear-Instituto Superior Técnico, Av.Rovisco Pais, 1049-001, Lisboa, Portugal*

**Abstract.** Local Alfvén wave (LAW) resonances are excited in tokamak plasmas by an externally driven electromagnetic field, with low frequency $\omega_A = c_A k_\parallel \ll \omega_{ci}$, where $c_A$ is the Alfvén velocity. Recently, wave driven density fluctuations at the LAW resonance $m=\pm 1$, $N=\pm 2$ with few kW power deposition and 4 MHz frequency were detected in TCABR ($B_t=1.1$T, $q_0=1.1$, $n_0=1.4$-$2.0\cdot 10^{19}$ m$^{-3}$) using a fixed frequency (32.4GHz) O-mode reflectometer. Here, we show that combination of small power deposition in LAW resonances, swept by plasma density variation or scanned with generator frequencies, in combination with detection of the density fluctuations in the LAW resonances by reflectometry can serve as diagnostic tool for identification of the effective ion mass number $A_{ef}$ and $q$-profile in tokamaks. The idea is based on the simultaneous detection of the position of $m=\pm 1$ local AW resonances, which are excited by M/N=$\pm 1/\pm 2$ antenna modes, and $m=0$ generated by poloidal mode coupling effect in tokamaks. According to LAW dispersion, the $m=0$ resonance depends only on the effective ion mass number and does not depend on the $q$-profile. Then, using these data we can define $q$-value at the position of $m=\pm 1$ L AW resonances. Using TAE coils with even toroidal modes ($N=2, 4,..$) and AW generator in the frequency band 0.6-1 MHz, an application of this method to Joint European Torus ($B_t=2.3$T, $q_0=1$-1.3, $n_0=5$-$7\cdot 10^{19}$ m$^{-3}$) is demonstrated.

The idea of strong resonant absorption of RF fields, which are excited at the local Alfvén wave (LAW) resonance in magnetically confined inhomogeneous plasmas, was based on a series of the theoretical works carried out at 70-80th (for example, [1-3]). LAW resonance excitation is defined by mode conversion of an externally driven RF field below the ion-cyclotron frequency, $\omega_{ci}$, into the kinetic or electrostatic Alfvén wave (shear AW) at the Alfvén resonance layer where the resonance density can be defined by the equation,

$$k_\parallel^2 = \frac{\omega^2}{c_{Ai}^2}\left[\frac{A_i n_i / n_e}{1-\omega^2/\omega_{ci}^2} + \frac{A_z n_z / n_e}{1-\omega^2/\omega_{cz}^2}\right], \quad k_\parallel = \frac{1}{R_0}\left(N + m/q(r)\right), \quad c_{Ai} = B_t \Big/ \sqrt{\mu_0 m_i n_e} \qquad (1)$$

where $m$ and $N$ are toroidal and poloidal wave numbers, $B_t$ is the toroidal magnetic field, $q$ is the safety factor and $A_z$ is the main impurity mass number. The eq (1) can be reduced to $\omega_A = c_A k_\parallel \ll \omega_{ci}$ for low frequency, where $c_A = c_{Ai}/\sqrt{A_{ef}}$ is the Alfvén velocity, $A_{ef}=(A_i n_i + A_z n_z)/n_e$ is effective mass number, and $\omega_{ci}$ is the ion cyclotron resonance frequency. The continuous spectrum (1) of the LAW resonances is known as the Alfvén wave continuum. In the standard quasi cylindrical model for wave excitation in tokamak plasmas, the oscillating RF field is represented as a sum of helical harmonics $\exp[i(m\theta+N\phi-\omega t)]$. These fields are excited in the plasma by an antenna current sheet, $J_{\theta,\phi}(r,\theta,\phi,t) = \Sigma_M J_{\theta,\phi}(M,N)\delta(r-b)\exp[i(M\theta+N\phi-\omega t)]$. The antenna spectra of $J_{\theta,\phi}(M,N)$ driven in tokamaks are very wide both for the toroidal and poloidal wave numbers; however, a few harmonics can satisfy the local AW resonance conditions (1). Due to toroidal effects, the poloidal wave numbers *m* in the plasma may be different from *M* that are excited by the antenna. Moreover, AW can also be excited in a plasma column as global Alfvén waves (GAW) [1-2], with wave fields corresponding to the discrete eigenfunctions of the proper boundary value problem. Toroidicity induced Alfven eigenmodes (TAE) as global waves can be excited at the position of bifurcation of the LAW resonances for different *m* and *N* numbers. The close position of GAW ($\Delta\omega_{GW} << \omega_{th}$) to the AW continuum minimum $\omega_{th}$, and its easy identification with magnetic probes together with the set of TAE resonances calculated from eq.(1) were explored as a diagnostic tool (named Magneto-Hydro-Dynamic diagnostics) for definition of the effective ion mass number $A_{ef}$ and *q*-profiles in a series of tokamak experiments (for example, [4-6]). However, the strong dependence of $\Delta\omega_{GW}$ on toroidal wave number and current and density profile produced substantial difficulties in the interpretation of the experimental data.

Recently, using a fixed frequency (32.4GHz) O-mode reflectometer, wave driven density fluctuations at the local Alfvén wave (LAW) resonance $m=\pm 1$, $N=2$, with frequency $f = 4$ MHz and relatively small power deposition of 30 kW, were detected in the Tokamak Chauffage Alfvén Brésilien (TCABR) [7]. Here, based upon results of calculations carried out with the ALTOK code [8], we propose to use this effect in JET.

**Plasma Model of TCABR and JET** The calculations with ALTOK code have been carried out assuming circular, in TCABR, and D-shape cross-section ($\kappa=1.43$, $\delta=0.25$) in JET. To obtain good accuracy, 473x99 mesh points are used in the code calculations. Here, we analyze the AW absorption in TCABR shot N$^o$10668 [7,9] (minor radius *a*=0.18m, major radius $R_0$=0.615m, toroidal magnetic field *B* = 1.15T, $I_p$ =78-80 kA, with safety factor in the center *q(0)* =1-1.1, line averaged plasma density $\underline{n}$ =1.2-1.5×10$^{19}$ m$^{-3}$, central electron and ion temperatures $T_{e0}$=450 eV and $T_{e0}$=150eV). The TCABR antenna module has two groups of RF current carrying straps, which are positioned in two toroidal cross-sections separated toroidally by an angle about 22$^o$, creating mainly the spectrum of the poloidal *M*= ±1, ±2 and toroidal *N*=±1, ±2, ±3… modes. Their local AW resonances are strongly separated by choosing 4 MHz generator frequency. In JET calculations, we use shot #62209 as a reference; minor radius *a*=1.05m, major radius $R_0$=2.85 m, toroidal magnetic field *B* = 2.3T, and for the following basic plasma parameters: current $I_p$ = 2.4 MA (safety factor *q(0)* =1.2), central plasma density $\underline{n}$ =6×10$^{19}$ m$^{-3}$ (gas deuterium), central electron and ion temperatures are is 4000 and 2500 eV, respectively. The generator frequency has been swept in the band *f*=0.6-1.0 MHz. The antenna module is proposed to have two groups of RF current carrying straps.

These groups are situated at two opposite toroidal positions creating mainly the spectrum of the poloidal $M= 0, \pm1, \pm2$ and toroidal $N=\pm2, \pm4$ modes. The plasma profiles used in the code calculations are quasi parabolic temperature profile $T=T_0(1-\Psi^{0,9})$, density profile of TCABR $n=n_0(1-\Psi^{0,7})$, rather flat profile of JET $n=n_0(1-\Psi^{0,4})$, and current profile is $j=j_0(1-\Psi^{0.9})^{1.2}$, where $\Psi$ is the normalized poloidal magnetic flux.

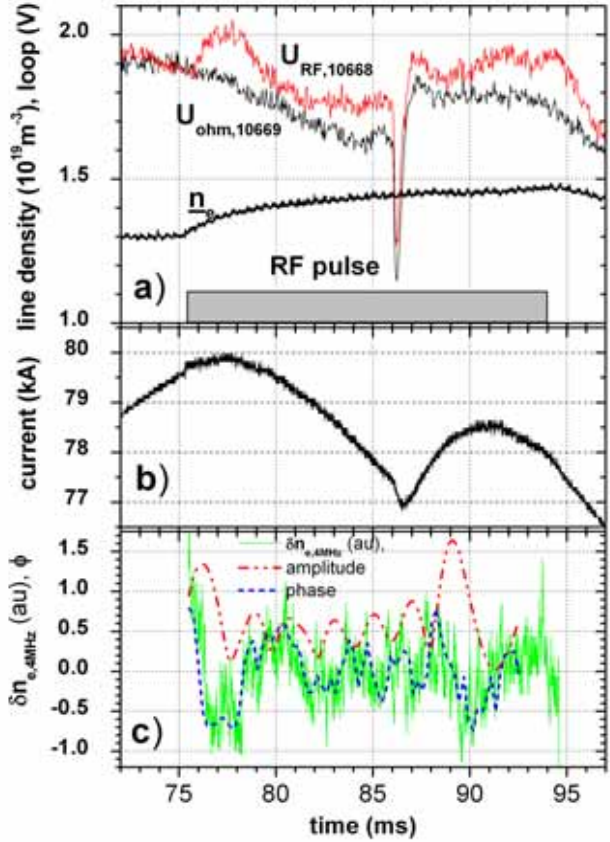

**Identification of Local AW Resonance with Reflectometry** The density fluctuation rise in the LAW position can be detected by a reflectometer. In the simple case, the amplitude of the density fluctuations is proportional to the parallel component of the electric field and to the square root of the absorbed power density, in accordance with Boltzmann distribution,

$$\frac{\delta n}{n_e} = \frac{e|\tilde{E}_\parallel|}{m_e k_\parallel v_{Te}^2}; \quad |\tilde{E}_\parallel| = 2\sqrt{\frac{\sqrt{8\pi}k_\parallel^3 v_{Te}^2}{\pi\omega_A^2 \omega_{pe}^2}}\tilde{p}$$

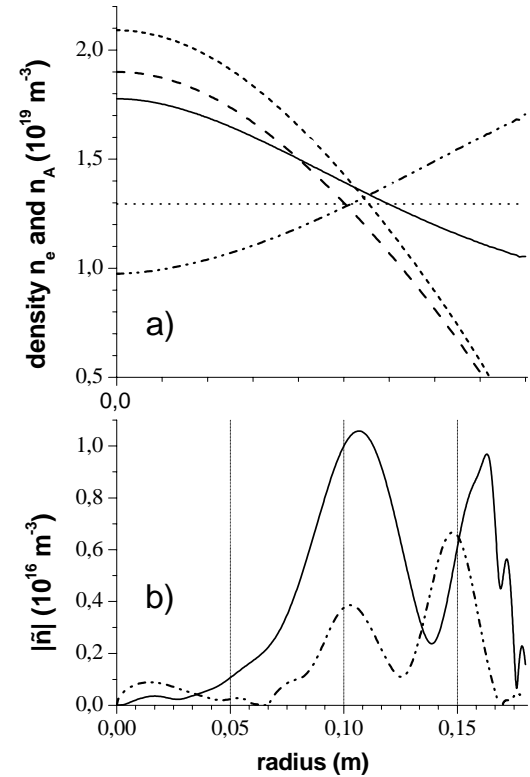

*Fig.1. Time traces (70-97 ms) of the loop voltage in comparison with Ohm discharge # 10669 and line averaged density (a), toroidal current (b) and respective amplitude of oscillating phase measured with reflectometer (c) in the TCABR discharge with AW heating (# 10668).*

*Fig.2. Plot of Alfvén density distributions $n_A$ (a) for $A_{ef}=1.1$, $q_0=1.07$, m/N=-1/3 mode (dash-dot-dot line), and for $A_{ef}=1.17$, $q=1.08$, m/N=-1/-2 (solid line) over radius, the electron density distributions for $\underline{n} = 1.3.\times10^{19}m^{-3}$ (dashed) and $\underline{n}=1.5\times10^{19}$ $m^{-3}$ short dashed lines), the cut off density is marked by dotted line; (b) the amplitudes of the*

*AW density fluctuations for 4 MHz frequency, respective to (a), m/N=-1/3 mode (dash -dot -dot line) and m/N=-1/-2 (solid line) resonances for $\underline{n}$ =1.5×10$^{19}$ m$^{-3}$.*

For 0.2kW/m$^3$ of power density at the resonance surface $r_A=0.65(R-R_0)$ (or about of 1kW of absorbed power), we obtain $\delta n/n_e \approx 1.10^{-4}$, which is of the level that can be detected by a modern reflectometer. The sensitivity of the system may be improved with locked frequency detection at the AW frequency. In Fig.1, we show the time traces (70-100 ms) of the loop voltage, line averaged density, toroidal current, and respective amplitude of 4MHz oscillating phase measured with the reflectometer in TCABR discharge # 10668, with AW heating. The perturbed density trace during RF pulse is restored from bolometry signal $\propto Z_{eff}^4 \, n^2/\, T^{3/2}$ assuming that there is small $Z_{eff}$ rise from 1.5 to 1.8, as found from ASTRA calculations, using the small loop voltage increase $Z_{eff}/T^{3/2}$ and $\beta_{eq}$ at the beginning of the RF pulse and diminishing later in comparison with a pure ohmic discharge # 10669. The central electron temperature is not increased during RF pulse, as indicated by the ECE signal. In Fig.1c, we can observe the maxima of the amplitude of 4 MHz oscillating microwave phase that should be interpreted as local AW resonances for $m/N$= -1/-2 ($A_{ef}$=1.17) in the initial stage and superposition with $m/N$= -1/3 ($A_{ef}$=1.1) in the final stage of the RF discharge # 10668.

**Alfvén wave Absorption and Fields in JET**

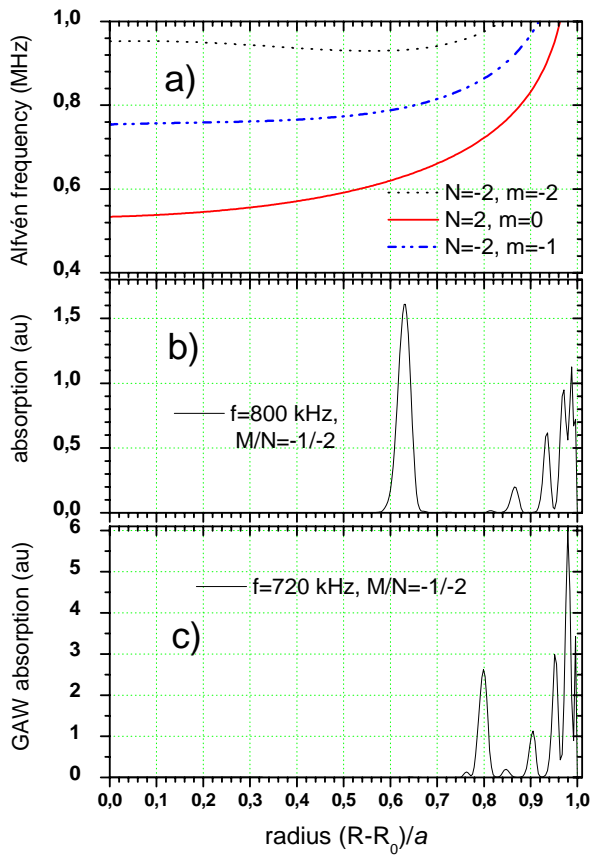

The AW continuum is shown in Fig.3a, where the AW frequencies for $m=0,\pm 1$, $N$=-2 are calculated with the respective "cylindrical" $q$-parameter

$$q_c = \frac{\kappa r B_\zeta}{R_0 B_\theta \sqrt{1-r^2/R_0^2}} \; .$$

The respective absorption profiles for 800 kHz and 720 kHz frequency for the $M$=-1, $N$=-2 antenna modes are presented in Fig. 3b,c. The absorption in Fig.3c is produced in the GAW resonance of the $m/N=-1/-2$ mode. There is good correlation of the local AW resonance position of the $N$=2 and $m$=0 in Fig.3a with the first absorption spike at $R-R_0= 0.8\,a$ in Fig.3c, calculated for the $m=0$ mode.

*Fig.3. Plot of distribution of AW continuum frequency for m=0,±1, N=2 (a), absorption profile for frequencies f=800kHz (b), and f=720kHz (c) in JET.*

A typical absorption in the *m/N=-1/-2* AW continuum is shown in Fig.3b for 800 kHz frequency. The corresponding power deposition profile, over the plasma cross section, is shown in Fig.4. The half width of those spikes are 3% of the plasma minor radius, i.e., smaller than 4cm and therefore offering good possibility for identification of the resonance position by reflectometry.

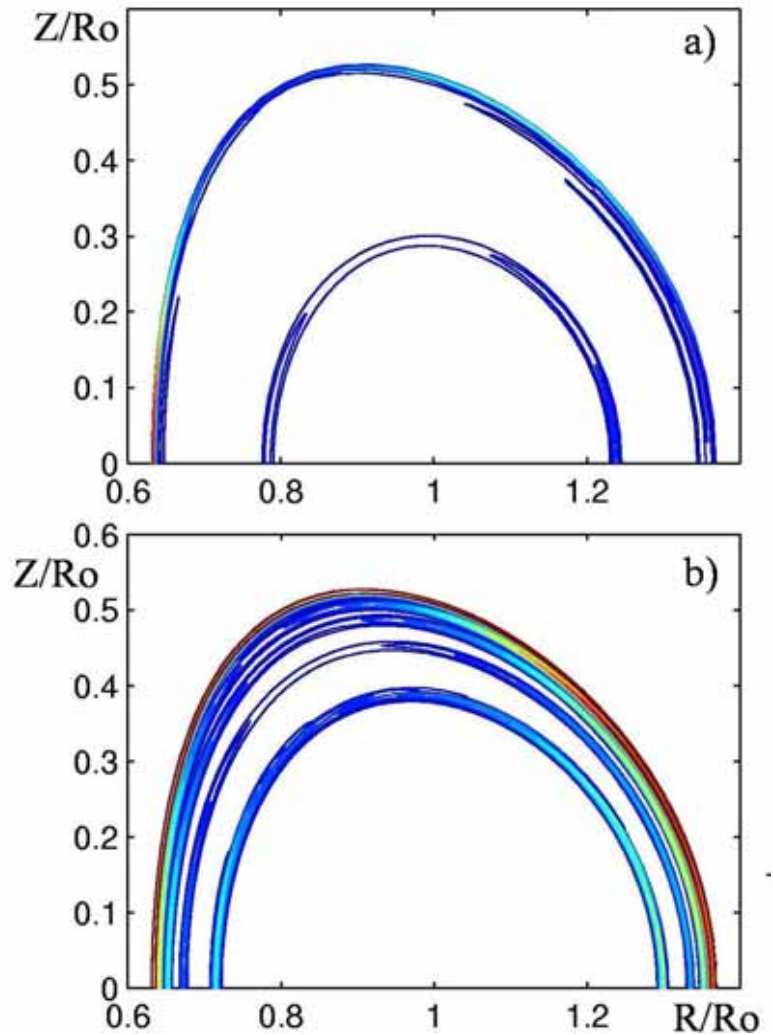

*Fig.4. Plot of the power deposition profile over the JET cross section ($Z/R_0$, $R/R_0$), corresponding to the case of Fig.3b and c.*

## Conclusion.

A combination of sweeping AW excitation system with the ECE radiometry and the sweeping of the reflectometer frequency can be a very powerful diagnostic tool to localize the AW power deposition, and to find effective mass number $A_{ef}$ and *q*-profiles in JET experiments.